\pdfoutput=1

\documentclass[11pt]{article}

\usepackage{ACL2023}
\usepackage{graphicx}
\usepackage{subcaption}
\usepackage{tabularx}
\usepackage{caption}
\usepackage{booktabs}

\usepackage{times}
\usepackage{latexsym}

\usepackage[T1]{fontenc}

\usepackage[utf8]{inputenc}

\usepackage{microtype}

\usepackage{inconsolata}

\title{TIMIT Speaker Profiling: A Comparison of Multi-task learning and
Single-task learning Approaches}

\vspace{0.9cm}
\author{Rong Wang* \\ University of Stuttgart, Germany\\
	\texttt{rong.wang@ims.uni-stuttgart.de}; \\ \texttt{rongw.de@gmail.com}\\  \\\And	
	Kun Sun \\
	University of Tübingen \\
	\texttt{kun.sun@uni-tuebingen.de}
}

\date{%
	\begin{minipage}{0.5\textwidth}
		\centering
		\texttt{rong.wang@ims.uni-stuttgart.de}
	\end{minipage}%
	\begin{minipage}{0.5\textwidth}
		\centering
		\texttt{kun.sun@uni-tuebingen.de} 
	\end{minipage}
}

\begin{document}
\maketitle
\begin{abstract}
This study employs deep learning techniques to explore four speaker profiling tasks on the TIMIT dataset, namely gender classification, accent classification, age estimation, and speaker identification, highlighting the potential and challenges of multi-task learning versus single-task models. The motivation for this research is twofold: firstly, to empirically assess the advantages and drawbacks of multi-task learning over single-task models in the context of speaker profiling; secondly, to emphasize the undiminished significance of skillful feature engineering for speaker recognition tasks. The findings reveal challenges in accent classification, and multi-task learning is found advantageous for tasks of similar complexity. Non-sequential features are favored for speaker recognition, but sequential ones can serve as starting points for complex models. The study underscores the necessity of meticulous experimentation and parameter tuning for deep learning models.
\end{abstract}
\pagestyle{plain}
\section{Introduction}
The TIMIT dataset \cite{garofolo1993darpa}, developed in 1986 by MIT in partnership with Texas Instruments, is a renowned corpus of continuous acoustic-phonetic speech. Featuring recordings from 630 speakers across eight US accent regions, each provides 10 phonetically rich utterances. Designed primarily for speech and speaker recognition research, it also supports applications like gender, age, and accent identification, facilitating the study of speaker profiling. In this study, we experiment with the TIMIT dataset for different speaker profiling tasks exploring different approaches, models, and feature sets. 
\par
Speaker recognition identifies a person by analyzing their voice characteristics based on seen speech data. For this task, We tried different feature sets to explore how feature engineering impacts the performance of models. We also include three other tasks namely gender classification, age estimation and accent classification. For these three tasks, We compared the differences between end-to-end multi-task learning and single-task learning approaches.
\par 

\section{Related work}
Researchers have made attempts to estimate speakers' physical attributes like height, weight, and age from speech data \cite{singh2016short}. Prior studies employed i-vector or x-vector, speaker embeddings created by CNN, for age estimation \cite{kwasny2020joint}, \cite{kaushik2021end}. Regional accent identification is less explored than gender and age prediction. \cite{shon2018convolutional} employed a CNN on features like MFCCs to identify Arabic accents, and most studies focused on native-speaker accent classification. \citep{lee2019gender} created an identification system for gender, age, and accent using separate deep learning models, later combined into a single system. To our knowledge, no present work addresses joint speaker profiling for age, gender, and accent.
\par
Multi-Task Learning (MTL) in machine learning seeks to improve generalization across multiple related tasks by harnessing shared information \cite{zhang2021survey}. The premise of MTL is that correlated tasks can share feature extraction, facilitating shared deep neural network layers. Yet, determining positive task correlations is challenging. \citep{tang2016multi} highlights tasks like language identification versus speaker recognition can be negatively correlated. Furthermore, \cite{marquet2023comparison} suggests MTL's strengths can also be its drawbacks. This ambiguity in task performance, when trained jointly versus separately, motivates our exploration of multi-task versus single-task learning.

\par
The advantages of multi-task learning over conventional single-task learning are still controversial. In this study, we want to provide experimental evidence for the benefits and weaknesses of multi-task learning in the context of gender, age and accent prediction when compared with single-task learning models. TIMIT dataset contains recordings from 630 speakers from eight accent regions of the United States, making it an ideal resource for accent recognition, besides age estimation, gender classification and speaker ID identification. While deep learning has minimized the necessity for manual feature engineering, skillful feature engineering remains essential for unlocking the full potential of deep learning models, leading to improved performance, generalization, and efficiency. This study will validate the importance of feature engineering for speech-related tasks through experimentation ion with different feature.

\section{Data Pre-processing}
TIMIT is provided with a standard division to Train and Test subsets. The Train subset consists of 462 speaker recordings (326 male and 136 female) and Test subset of 168 speaker recordings (112 male and 56 female). In both subsets, each speaker utters 10 sentences. Speakers cover seven major dialact regions plus a set of speakers moving around.
\begin{figure}[h]
    \centering
    \includegraphics[width=0.5\textwidth]{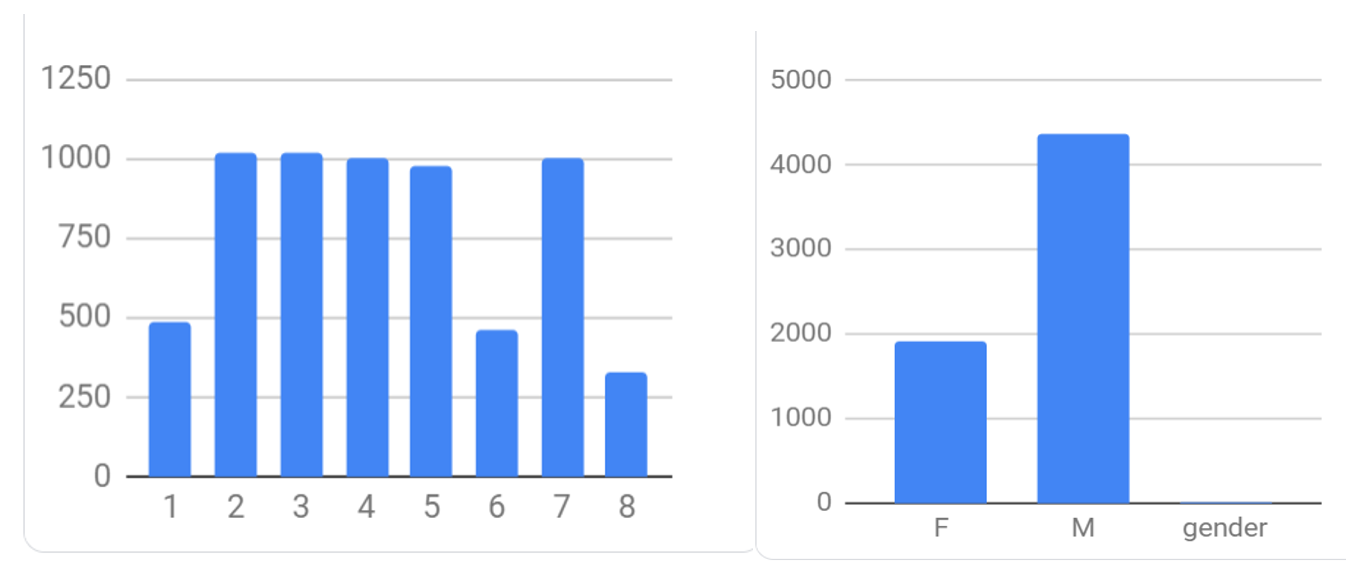}
    \caption{Imbalanced labels of Accent and Gender}
\end{figure}
\par
As Figure 1 shows, TIMIT dataset is highly unbalanced in terms of gender and accent distribution, which may lead to biased learning, where the model might perform well in the majority class while struggling with the minority classes. In order to balance both gender and accent classes, a combined label (accent\_gender) in the df column was created to oversample the two classes together. This operation results in data augmentation, increasing the training data from 4610 to 9440 samples.
In the context of speaker identification task, we combine the train and test dataset to split it into train,val and test dataset, no need of data balancing. 

For the TIMIT dataset, with inherent attributes like age, gender, and accent for each speaker, there's a risk that models might overfit by memorizing speaker-specific traits instead of genuinely predicting these attributes. Although using StandardScaler() can help with speaker normalization, it might not handle complex feature combinations well. Methods like Z-score and Cepstral Mean Normalization (CMN) could be more effective but were not tested due to time constraints.

In this study, the primary challenge was creating a TIMIT class (Dataset) to extract features from raw audio, encode diverse labels, and align features with their labels. A data frame was proved to be more suitable than a dictionary for storing the data, and the \_getitem\_ function was designed to simultaneously provide all necessary inputs like features, feature length, speaker id, gender, accent and age for forthcoming trainings.

There are generally two types of speaker normalization methods: across-Speaker Normalization and speaker-wise normalization. Across-speaker normalization involves calculating statistical measures, such as mean and standard deviation, across all speakers' data and aims to remove speaker-specific variations, allowing the model to focus on the shared characteristics among speakers.  normalizing each speaker's data using these global statistics, while speaker-wise normalization involves normalizing each speaker's data individually. In the context of age, gender, and accent recognition, across-speaker normalization can be particularly useful. 

\section{Gender, Age and Accent Prediction: Multi-Task vs Single-Task}

Single-task models are straightforward, immune to negative transfer risks since they focus on one task, and particularly advantageous in specialized domains. Here, we detail experiments aiming to optimize performance across three specific tasks, providing a benchmark for assessing multi-task learning models.

\subsection{Single-Task Learning}
\textbf{Gender classification} from audio signals is simple due to distinct acoustic traits between genders. Averaging MFCCs over time and expanding MFCCs from 13 to 40, the performance of a three-layer feed-forward network is shown in table 1.
\begin{table}[h]
\centering
\caption{Gender Classification with Different MFCCs}
\small
\begin{tabular}{@{}lccccc@{}}
\toprule
 & MFCCs & Accuracy & Precision & Recall\\
\midrule

&13 & 0.941 & 0.941 & 0.941 \\
 &30 &  0.986 & 0.986 & 0.986 \\
 &40 &  0.986 & 0.986 & 0.986 \\

\bottomrule
\end{tabular}
\end{table}

\par
\textbf {Accent classification} is trickier than other audio classification tasks due to its nuanced pronunciation differences, unlike the clear acoustic variations in gender. Accents are also context-dependent and subjective\cite{zhang2021accent}. This study faced challenges in accent recognition, experimented with various features and models, and presents results in Table 2. Five Types Features refer to mean MFCCs, Mel, Chroma, Tonnetz and Contrast.
\begin{table}[h]
\centering
\caption{Accent Classification Results.}
\small
\begin{tabularx}{\linewidth}{@{}lXXX@{}}
\toprule
Feature-Model Configuration & Accuracy & Precision & Recall\\
\midrule
Sequential MFCC(30)-CNN & 0.10 & 0.13 & 0.11 \\
Sequential MFCC(40)-LSTM & 0.16 & \textbf{0.18} & 0.16 \\
MFCC(40)-MLP & 0.18 & 0.16 & 0.18 \\
Five Types Features-MLP & \textbf{0.21} & 0.16 & \textbf{0.21} \\
\bottomrule
\end{tabularx}
\end{table}

\textbf{Age estimation} using the TIMIT dataset is intriguing, though the age distribution is imbalanced. We explored MLP, LSTM and CNN structures to do the regressional task. The experimental results are shown in Table 3.
\begin{table}[h]
\centering
\caption{Age Estimation with different models}
\small
\begin{tabular}{@{}lccccc@{}}
\toprule
Feature-Model Configuration & MAE & RMSE\\
\midrule
MFCC(40)-MLP & 6.16 &10.82\\
sequential MFCC(30)-LSTM & 6.02 & 10.26\\  
sequential MFCC(30)-CNN & \textbf{5.53}  & \textbf{9.24}\\
\bottomrule
\end{tabular}
\end{table}
\par
It is clear that CNN model works best for age estimation task compared with LSTM and MLP models. The CNN model employs two convolutional layers(kernel size 3, stride 1, and padding 1) with increasing channels, followed by ReLU activations and max pooling (kernel size 2, stride=2) for hierarchical feature extraction from sequential MFCCs (30) features.

\textbf{Discussion} Single-task learning offers advantages in specific scenarios, yet achieving all-encompassing models for all three tasks seems unattainable. The above experimental results shows that each task benefits from unique architectures: a 3-layer MLP, a 2-layer LSTM, and a 2-layer CNN. While gender classification proves versatile with various feature-model combinations, age estimation excels with CNN using sequential MFCCs, and accent classification favors MLP with non-sequential features. 

\par

\subsection{Multi-Task Learning}
To explore MTL, we experimented with different model architectures, MultiTask MLP and MultiTask CNN+LSTM Model. The first model, MultiTask MLP, is a traditional feed-forward network designed for multi-task learning on non-sequential feature data. The second model, MultiTask CNN+LSTM, is more complex, integrating convolutional and recurrent layers, making it suitable for processing sequential audio data like MFCCs for three tasks: age prediction, gender classification, and accent classification. 
\par
\textbf{MultiTask CNN+LSTM Model} is optimized for processing MFCC features varying from 13 to 40 coefficients. The shared layers have two 1D convolutional layers with max-pooling for feature extraction from audio sequences, followed by an LSTM layer for capturing temporal dependencies so that the model extracts both spectral and temporal patterns from audio. The task-specific layers have feed-forward networks for accent (8 classes) and gender (2 classes) classification, and a linear layer for age prediction. 
Table 4 shows the results of MultiTask CNN+LSTM Model using sequential MFCC features varying from 13 to 40.
\begin{table}[h]
\centering
\caption{Performance of MultiTask CNN+LSTM Model using different numbers of sequential MFCC features}
\small
\begin{tabular}{@{}lccccc@{}}
\toprule
 & MFCC & Task & Accuracy & Precision & Recall\\
\midrule

 &13 & Age & MAE 6.03 \\
   & & Gender & 0.98 & 0.98 & 0.94 \\
   & & Accent & 0.14 & 0.14 & 0.14 \\
 &25 & Age  & \textbf{MAE 5.97} \\
    & & Gender & 0.98 & 0.99 & 0.96 \\
    && Accent &\textbf{0.15} & \textbf{0.14} & \textbf{0.14} \\
 &40 & Age & MAE 6.08 \\
    & & Gender & \textbf{0.99} & \textbf{0.99} & \textbf{0.96} \\
    && Accent & 0.11 & 0.11 & 0.11 \\

\bottomrule
\end{tabular}
\end{table}

\subsection{Single-Task vs Multi-Task with MLP}

To enable meaningful comparison, we experimented with 3 layer-MLP architecture for both single-task learning (STL) and MTL with consistent parameters, the same model and hyperparameters. The only difference lies in the implementation method. In MLT, three tasks were trained and predicted simultaneously, while in STL, three single-task models were trained and optimized individually. Due to high computational costs, we didn't test CNN+LSTM for single task learning.

\textbf{MultiTask MLP Model} utilizes a 1-dimensional (averaging features across time frames) variable feature combination of MFCC, Mel, Chroma, Tonnetz and Contrast, designed to simultaneously predict accent, gender, and age. The model has three shared linear layers interspersed with ReLU activations, dropout, and batch normalization then followed by task-specific layers composed of linear layers for accent classification (8 classes), gender classification (2 classes), and age prediction (regression).
\par
Table 5 details the MLP performance using MFCC+Mel features, Table 6 presents results with MFCC+Mel+Chroma+Tonnetz+Contrast features.  
\begin{table}[h]
\centering
\caption{STL vs. MTL with MFCC+Mel}
\small
\begin{tabular}{@{}lccccc@{}}
\toprule
 & Model & Tasks & Accuracy & Precision& Recall\\
\midrule

 MTL &MLP & Accent & 0.13 & 0.14 & 0.13 \\
      & & Gender & 0.97 & 0.97 &0.97\\
      & & Age  & MAE 7.71  \\
 STL & MLP & Accent & \textbf{0.16} & \textbf{0.16}& 0.16   \\
     & & Gender & \textbf{0.98} & \textbf{0.98}& 0.98\\
       & & Age  & \textbf{MAE 6.66} \\ 

\bottomrule
\end{tabular}
\end{table}

\begin{table}[h]
\centering
\caption {STL vs. MTL with five types of features}
\small
\begin{tabular}{@{}lccccc@{}}
\toprule
 & Model & Tasks & Accuracy & Precision &Recall\\
\midrule

 MTL &MLP & Accent & 0.12 & 0.15 & 0.12 \\
      & & Gender & 0.97 & 0.97 &0.97\\
      & & Age  & \textbf{MAE 6.17}\\ 
 STL & MLP & Accent & \textbf{0.15} & \textbf{0.21} &0.15  \\
     & & Gender & \textbf{0.99} & \textbf{0.99} &0.99\\
       & & Age  & MAE 6.18  \\

\bottomrule
\end{tabular}
\end{table}

\subsection{Discussion}
Multi-task learning, through a unified model, seeks to improve the performance of multiple related tasks. While our findings indicate that multi-task models deliver performances on par with single-task models in gender and age estimation, they fall short in accent classification. This difference in performance can be attributed to the varying complexities of each task. Gender classification, for example, is relatively straightforward, but accent classification becomes challenging due to the nuances in speaker accents. To address these complexity disparities, we experimented with weighted loss. For instance, when adjusting the loss weights to (5 * accent\_loss + gender\_loss + 0.01 * age\_loss), our MLP model saw a 3\% boost in accent prediction accuracy, but the age estimation's MAE deteriorated to 7.3. Conversely, using a weighted loss of (accent\_loss + gender\_loss + 0.001 * age\_loss), the overall performance of the CNN+LSTM model improved. These experiments underscore the intricacies of balancing multi-task losses, emphasizing the need for careful optimization. 

\section{Speaker Identification vs. Accent Recognition}
\par
\subsection{Challenges of Accent Recognition}
Our results highlight that accent recognition, with an accuracy below 25\% in both single and multi-task models, is especially challenging. This difficulty arises from the complexities in differentiating 8 American English accents compared to simpler binary classification tasks like distinguishing American from British accents. The model must differentiate among 8 various accents, emphasizing general accent patterns and shared dialectal traits and disregarding speaker-specific nuances. The task is stark contrast to speaker identification tasks, where the model learns to recognize speakers it has encountered before, thereby incorporating their unique characteristics.

In contrast, speaker identification models can benefit from recognizing voices from training, simplifying generalization. A key difference between speaker identification and age/accent/gender recognition is the presence of speakers in the training data. To emphasize the contrast in difficulty between these tasks, we further examined speaker ID recognition using different feature sets.
\subsection{Speaker Identification Task}
The goal of speaker identification is to determine the exact identity of the speaker among a set of known speakers based on his/her unique vocal characteristics. During the testing phase, the trained model takes in a new voice sample and attempts to match it with one of the known speakers in its training data. Apparently, the default train, test spilt of TIMIT is not applicable for speaker identification task, because speaker never overlap in train and test.The TIMIT corpus comprises 630 speakers, but due to one speaker having missing data, we worked with 629 speakers for classification, encoding their speaker IDs as labels.Given the non-overlapping nature of TIMIT's default train and test speaker sets, we combined and split the data into 70\% training, 10\% validation, and 20\% testing. 

\begin{table}[h]
\centering
\caption{Speaker ID Recognition Results}
\small
\begin{tabular}{@{}lccccc@{}}

\toprule
Features & Model & F1 Macro  & Precision & Recall\\
\midrule
MFCC(40)& MLP & 0.75  & 0.83& 0.79 \\
MFCC(40)& LSTM & 0.76  & 0.86& 0.80 \\
MFCC(40)+Mel(64)& MLP & 0.80 & 0.89 & 0.84  \\
MFCC(40)+Mel(64)& LSTM & 0.83 & 0.91  & 0.86\\
five types features& MLP & 0.80 & 0.88  & 0.84\\ 
five types features& LSTM & 0.83  & 0.91 & 0.86\\ 

\bottomrule
\end{tabular}
\end{table}

\textbf{Discussion} Despite the challenging conditions of just 10 samples per speaker and a daunting 629 classes (one speaker discarded due to missing data), the model's performance exceeds that of accent recognition, surpassing the 21\% in accent recognition using the same features and model. This demonstrates the model's ability in internalizing acoustic features, enabling it to recognize the speaker when presented with new audio samples. Interestingly, the 2-layer LSTM model outperformed the 4-layer MLP after 100 epochs with one-dimensional features, showing that more feature types do not necessarily lead to better performance, since five types of features have the same result as that of two.

\section{Conclusion}
This study investigated single-task and multi-task deep learning approaches across four speaker profiling tasks: gender classification, accent classification, age estimation, and speaker identification. The challenges of accent classification were evident. Here are insights addressing our initial research questions:
\par
\textbf{Multi-Task vs Single-Task}:
Comparing STL with MLT across all potential model structures is unfeasible. Our experiments with MLP demonstrate that although multi-task learning slightly improve age estimation, it compromises accent prediction. The findings confirm the previous research that MLT is best suited for related and similar complexity tasks. Joint prediction is proved to be more effective for tasks with similar complexity, such as age and gender \cite{kwasny2020joint}. The study suggests that multi-task learning benefits age estimation, but compromises the performance of accent prediction. It is proved that multi-task learning is best suited for related and similar complexity tasks, and the joint prediction of age and gender was proved to be advantageous by previous research \cite{kwasny2020joint} Given the extreme difficulty of accent recognition with the TIMIT dataset, the simultaneous integration of gender, age, and accent tasks may not be a good idea. Lack of correlation among the three tasks hinders their mutual enhancement, particularly with accent recognition.
\par
\textbf{Feature Selection}:
Unlike in automatic speech recognition(ASR), our findings suggest that speaker recognition tasks benefit from feature without time frames. Averaging MFCCs over frames counters the variability caused by short-term fluctuations, resulting in a more stable and higher-level representation, robust to audio length variations. One trick in feature extraction is to reduce the value of hop\_length, we use 160 instead of default 512 to extract all kinds of features in this study, as a smaller hop\_length value results in a larger number of frames to average. While sequential features didn't notably improve accent classification or speaker identification with MLP models, they're recommended as a starting point with CNN or LSTM models to find the hierarchical features.
\par
\textbf{Hyperparameters Tuning}:
The study focused on conventional models (DNN, CNN, LSTM) rather than complex ones such as Transformers. Hyperparameter tuning proved crucial with the conventional deep learning models. Factors such as batch normalization, dropout, and optimizer selection significantly influenced model performance, emphasizing the need for fine-tuning model configurations.
\par
In conclusion, these findings underscore the complex nature of deep learning models and the importance of meticulous experimentation and tuning. It's crucial to empirically validate assumptions about feature selection and model complexity through thorough testing.
\section*{Limitations and Future Work}
Due to limitations in time and computational resources, our study did not explore cutting-edge techniques like transfer learning, transformers-based model, or speaker-embedding method for improved speaker recognition. It is worth noting that Multi Task Learning may outperform Single Task Learning with other complex and intricate models to realize the full potential of Multi Task Learning in speaker profiling task. Future work could focus on accent recognition or jointly predicting gender, age, and height on TIMIT dataset.
\bibliography{reference}
\bibliographystyle{acl_natbib}
\end{document}